\def\cm2{cm$^{-2}$}

\def\c2{C~{\sc ii}}
\def\c4{C~{\sc iv}}
\def\fe2{Fe~{\sc ii}}
\def\fe3{Fe~{\sc iii}}
\def\mg1{Mg~{\sc i}}
\def\mg2{Mg~{\sc ii}}
\def\si2{Si~{\sc ii}}
\def\si4{Si~{\sc iv}}
\def\al2{Al~{\sc ii}}
\def\al3{Al~{\sc iii}}
\def\o1{O~{\sc i}}
\def\n1{N~{\sc i}}
\def\h1{H~{\sc i}}

\def\approxlt{\mathrel{\spose{\lower 3pt\hbox{$\sim$}}
        \raise 2.0pt\hbox{$<$}}}
\def\approxgt{\mathrel{\spose{\lower 3pt\hbox{$\sim$}}
        \raise 2.0pt\hbox{$>$}}}

\documentclass[article]{has40}
\usepackage{graphicx}
\usepackage{amssymb}
\tabletypesize{\normalsize}  

\def\plotone#1{\centering \leavevmode
\includegraphics[width=.95\columnwidth]{#1}}

\def\plotone#1{\centering \leavevmode
\includegraphics[width=.95\columnwidth]{#1}}

\shortauthors{Stellingwerf}
\shorttitle{Three Dimensional Models}

\begin{document}
\large    
\pagenumbering{arabic}
\setcounter{page}{28}

\title{Three Dimensional Models of RR Lyrae Pulsation}

\author{{\noindent R. F. Stellingwerf{$^{\rm 1}$}}\\
\\
{\it (1) Stellingwerf Consulting, 11033 Mathis Mtn Rd SE, Huntsville, AL 35803, USA\\} 
}

\email{(1) rfs@stellingwerf.com }

\begin{abstract}
Preliminary polytropic models of a pulsating star have been constructed using the 3-D Hydrocode SPHC. An embedded Rayleigh-Taylor unstable layer is used to study the interaction of pulsation and turbulence. The possible importance of Richtmyer-Meshkov instability is noted for large amplitude pulsation. In addition, this model spontaneously generates a strong, hot, shock driven outflow at the upper boundary.
\end{abstract}

\section{Introduction}
Stellar pulsation models have been remarkably successful in reproducing many of the features seen in pulsating stars, particularly RR Lyrae stars and Cepheids. We note that there are several areas that have not been fully modeled, including, 1) turbulence, 2) the outer boundary condition, 3) magnetic field effects, and 4) rotation. This memo describes some initial tests addressing the first two of these concerns using the three dimensional hydrodynamic technique of Smooth Particle Hydrodynamics.

\section{SPHC Turbulence Models}

SPHC is a "Smoothed Particle Hydrodynamics" (SPH) code (Lucy 1977, Monaghan 1982). The SPH technique is a grridless method of representing a continuous fluid by means of fixed-mass SPH "particles", each of which consists of a mathematical basis function (or kernel) that moves with the fluid. These kernels are cubic B-Spline functions that resemble a Gaussian in shape and have a half-width given by the smoothing-length (h). The kernels form a set of basis functions for all fluid quantities consisting of overlapping spheres of diameter 2*h, with weighting function for each particle's contribution at a given point in the vicinity given by the local value of the kernel. The integral of each kernel is normalized to unity, so the sum of the particle mass times the kernel can be visualized as a localized, but slightly fuzzy fixed-mass particle. Fluid quantities such as density and velocity are represented by sums of these basis functions, which are easily differentiated to compute the needed terms in the conservation equations of fluid flow. 

SPHC was first developed in 1987, and has been extensively tested and validated on many shock and flow problems (Stellingwerf 1990). See http://www.stellingwerf.com for background and examples.

The configuration of SPHC used in these models consists of a standard SPH approach (Monaghan 1988) using a Monaghan viscosity term (Monaghan 2005) to handle shock waves, with a  Balsara weighting on this viscosity to reduce shear dissipation (Balsara 1995). Particle size (h) is varied in these simulations in regions of varying density to maintain contact between adjacent particles. This automatically increases the resolution in regions of higher density, a desirable feature. The equation of state used is that of an ideal gas with the mean molecular weight and the ratio of specific heats ("gamma") chosen to match the desired cases. A small amount of conservative smoothing is applied to the thermal energy to avoid spikes caused by rapid shock heating. This correction is similar to, but much weaker than, the "wall-heating" thermal conduction correction proposed by Monaghan (Monaghan, Gingold 1983).

In stellar pulsation modeling hydrodynamic mixing and turbulence are of interest. Three types of instabilities are encountered: 1) Rayleigh-Taylor (RT), in which a denser fluid overlies a lighter fluid in a gravitational (or acceleration) field, 2) Kelvin-Helmholtz (KH), encountered at a shear interface, and 3) Richtmyer-Meshkov Instability (RMI), caused by a shock wave encountering density inhomogeneity. Since shocks are present in pulsating stars, it is likely that RMI plays a role. To clarify this process, we present an example.

SPHC has recently been tested on models of Richtmyer-Meshkov Instability in which a strong shock wave encounters density inhomogeneity (Stanic, et. al. 2012). The result is turbulent mixing induced by the passage of the shock. An example of such a model is shown in Figure 1 in which a strong shock encounters a density jump with a sinusoidal interface. Here the left boundary is a piston that follows the post shock flow. In more realistic cases and at later times the mixing region becomes completely turbulent.  In these test sequences SPHC results compared favorably to theory, to other code results, and to experimental results. RMI is expected in any fluid subject to shocks and containing the possibility of any sort of density inhomogeneity.

\begin{figure*}
\centering
\plotone{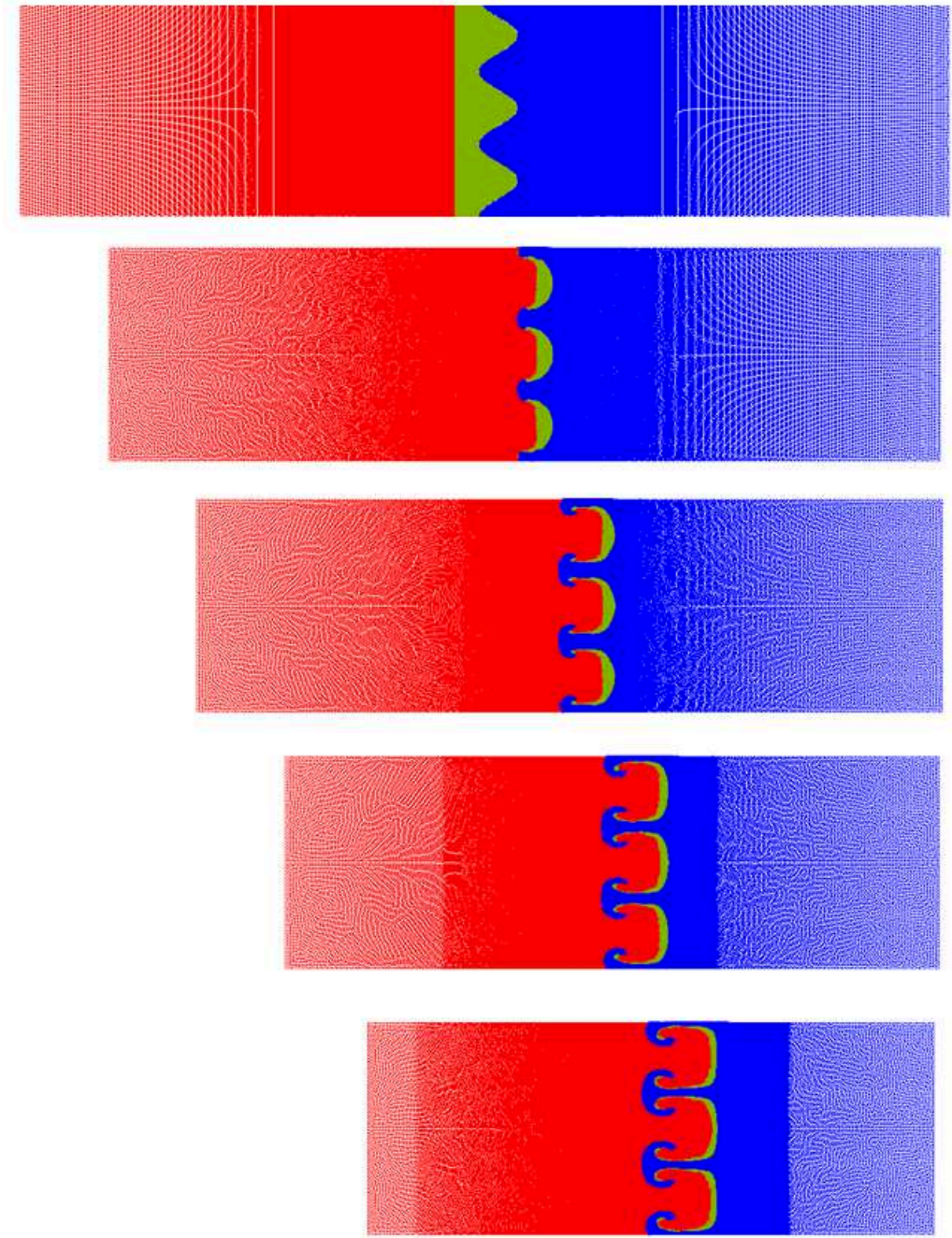}
\vskip0pt
\caption{Example of Richtmyer-Meshkov Instability. A strong shock (red/green interface in top panel) moving to the right, moves from a low density region (red) to a high density region (blue) across a perturbed interface. Subsequent times are shown from top to bottom. }
\label{o1039}
\end{figure*}

\pagebreak 

SPHC has also been successfully tested on cases of Rayleigh-Taylor instability and Kelvin Helmholtz instability. KH is often found embedded in other types of instabilities whenever shear layers are formed during a mixing event. We now gradually introduce pulsation into the test cases and move toward a stellar model in which 3-D turbulence plays a role.

\section{RR Lyrae Partial Envelope Model}

Pulsating stars have a region of convective instability in the envelope near the ionization regions of hydrogen and helium. This is a source of turbulent mixing that is modeled in some pulsation codes. It is believed (and usually found) that the level of turbulence is generally rather low in the instability strip (Stellingwerf 1982), although it is also believed that the importance of convection increases for cool RR Lyrae and Cepheid stars, eventually stabilizing the pulsational instability at the red edge. This unstable region can induce a density inversion, and hence a Rayleigh-Taylor instability.

A highly uncertain aspect of the convective RR Lyrae models is the interaction between convection and pulsation, especially for large amplitude pulsation in which strong shocks are present. To investigate this phenomenon, as well as test the SPHC code on this type of problem, a simple pulsation/turbulence model was constructed. This model consists of a cube filled with perfect gas, 0.25 solar radii on a side (about 5\% of an RR Lyrae star), with a gravitational field appropriate to an RR Lyrae model. The density of the gas is initially set twice as large in the top half of the cube, so the setup is Rayleigh-Taylor unstable, but the temperature is adjusted to maintain constant pressure throughout the cube at time zero. All boundaries are reflecting.

Initial motion of the model is downward as influenced by gravity. This induces a series of pulsational oscillations. Mixing at the unstable interface begins during the first expansion phase and continues rapidly thereafter. This model was run in two dimensions and in three dimensions. The result of the two models (colored on initial Y value) is shown in figure 2.

\begin{figure*}
\centering
\plotone{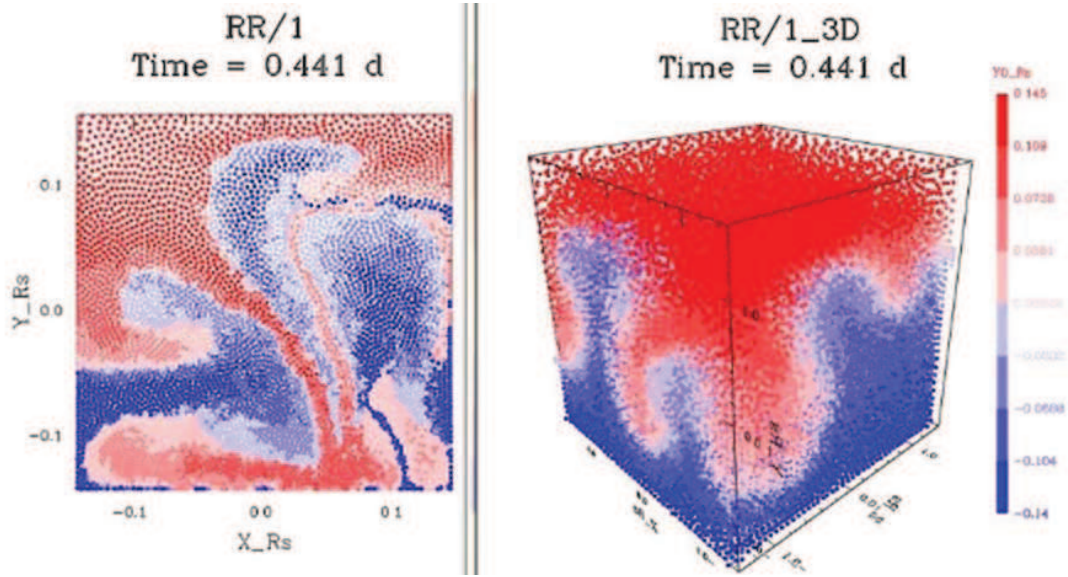}
\vskip0pt
\caption{Mixing phase of the Rayleigh-Taylor plus pulsation model described in the text. Left: 2-D model, right: 3-D model. }
\label{o1039}
\end{figure*}

An animation showing the full motion of this model can be viewed here:

{\noindent \small \bf http://www.stellingwerf.com/rfs-bin/index.cgi?action=GetDoc\&id=132\&filenum=1}

The two dimensional case forms a single eddy which overturns on a timescale of several pulsation periods. The mixing is turbulent and the result is a nearly complete inversion of the initially upper (red) and lower (blue) regions.

The three dimensional model is more complex. About 5 or 6 areas of upward flow develop and interact when they arrive near the top of the cube. The turbulence resembles that of the 2-D case, but no obvious eddys form in the 3-D flow, and the mixing is slower than the 2-D. 

In both cases, the mixing clearly accelerates during the expansion phase of the motion. This is interpreted as the effect of the shock that forms at maximum compression and subsequently passes through the unstable region during the following expansion. This suggests that Richtmyer-Meshkov Instability may be important in stellar pulsation, an effect not considered in existing models.

\section{RR Lyrae Full Sphere Model}

The model discussed in the previous section was then extended to spherical geometry. The main features of this model are shown in figure 3. The model is 4 solar radii in extent and is meant to model RR Lyrae itself. In this case the initial structure is that of a stellar polytrope with varying temperature and density to start with a condition in equilibrium with the inward radial gravitational field. The outer edge of the gas cloud has a positive pressure at time zero and is free to expand. In this particular case the masses of the outer SPH particles are sufficiently small to allow an outward expansion and extended atmosphere. Pulsation is initiated in this model by applying an initial inward velocity field increasing linearly from 0 to -20 km/s at the surface.

\begin{figure*}
\centering
\plotone{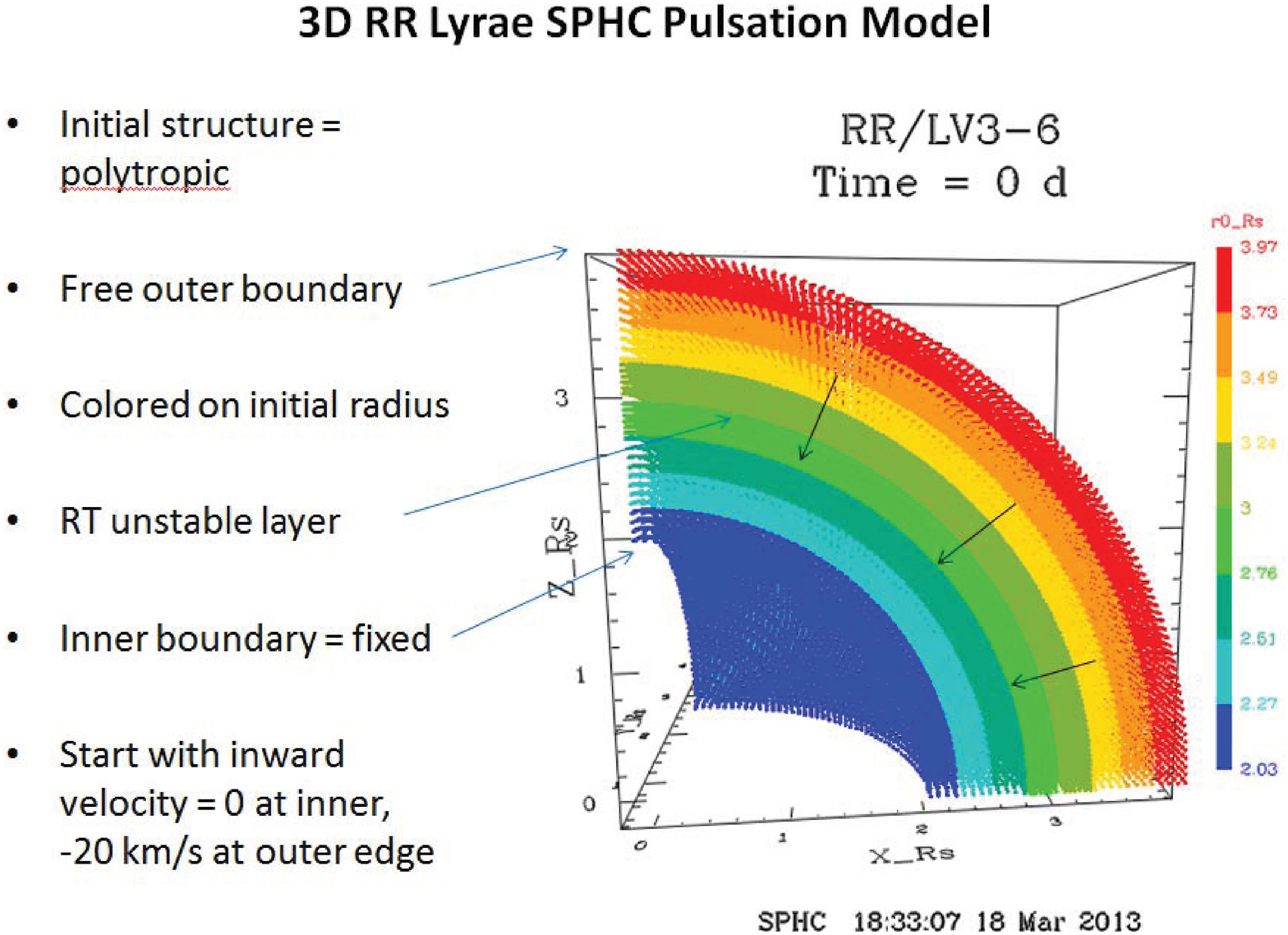}
\vskip0pt
\caption{Graphic showing the main features of the spherical pulsation/turbulence model.}
\label{o1039}
\end{figure*}

\subsection{Turbulent Mixing}

To study the effects of turbulent mixing in this geometry, a density inversion in pressure equilibrium was inserted in the initial model at the blue/orange interface. Figure 4 shows the extent of the mixing after several pulsation periods. 

\begin{figure*}
\centering
\includegraphics[width=9cm]{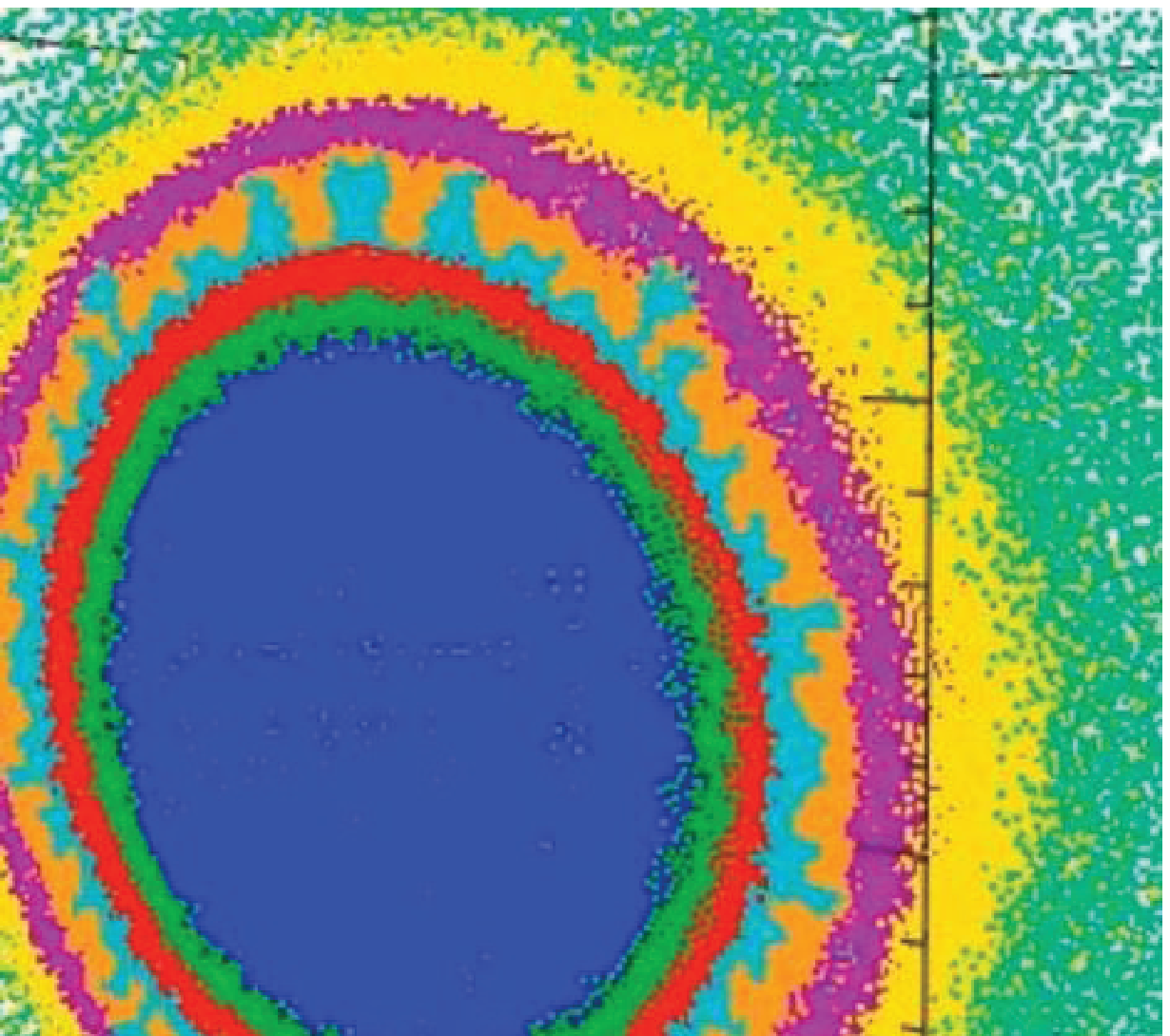}
\vskip0pt
\caption{Graphic showing the main features of the spherical pulsation/turbulence model.}
\label{o1039}
\end{figure*}

An animation showing the full motion of this model can be viewed here:

{\noindent \small \bf http://www.stellingwerf.com/rfs-bin/index.cgi?action=GetDoc\&id=133\&filenum=1}

Again, the mixing is observed to accelerate during the expansion phase of the pulsation, so Richtmyer Meshkov could be affecting the flow. One striking feature of this model is that substantial overshooting beyond the initially unstable layers is observed both above (into the purple layer) and below (into the red layer). We emphasize that his model does not include radiation, so the instability is not convective, as in actual stars. It is intended to approximate convective motion, but only for a few periods at most.

\subsection{Atmosphere/Wind Development}

As mentioned above, the zoning at the outer edge of the initial model was chosen to allow expansion of the outer layers if the physics favored such motion. In this case, substantial outward flow did occur. The outward expansion was initiated by the first shock to arrive from the pulsation motion. This propelled material out to about ten solar radii (2.5 stellar radii). This material then began to fall back into the star, to be met with the next shock-propelled outflow event. After a few periods the outer structure of the star settled down to a dynamic situation depicted in figure 5. A series of very strong, hot, turbulent shock waves move through material already present to form a structure resembling a corona. The blue dots in figure 5 represent 10,000 K, and the red dots in figure 5 represent temperatures of one-half million degrees K. In a real star this high temperature would be tempered by radiative cooling, but would still be very hot. Beyond about 10 solar radii (outer limit of plot) the material no longer falls back, but forms an outflowing wind. 

\begin{figure*}
\centering
\includegraphics[width=11cm]{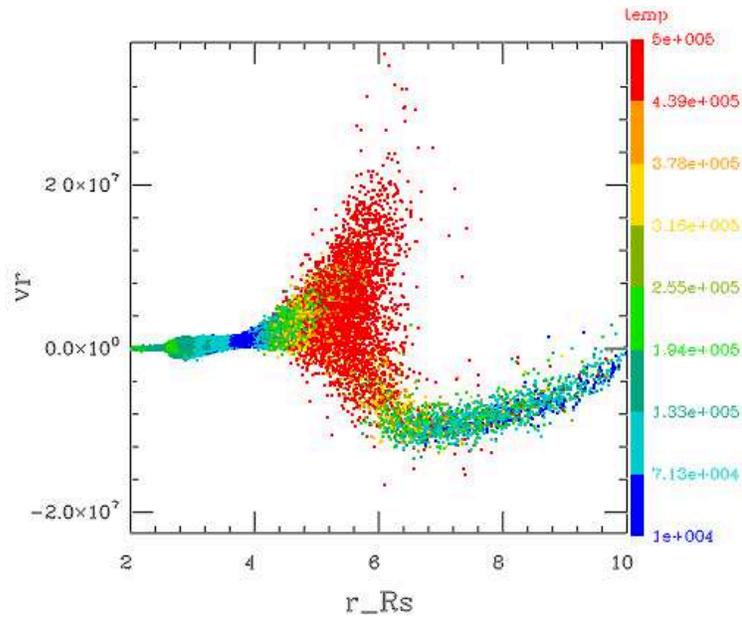}
\vskip0pt
\caption{Radial velocity versus radius, colored on temperature as the third pulsational shock begins to interact with infalling material}
\label{o1039}
\end{figure*}

An animation showing the full motion of this model can be viewed here:

{\noindent \small \bf http://www.stellingwerf.com/rfs-bin/index.cgi?action=GetDoc\&id=134\&filenum=1}

\subsection{4.3	Sky View}

This model can be optimized to provide an interpretation of the star's appearance to an outside observer sufficiently near to resolve the features seen here. To do this the above model was displayed with black background, in non-cutaway view, and with the color scale reversed. So, in this plot red represents 10,000 K, and blue represents one-half million degrees. Figure 6 shows one frame from this movie. Note the distinctly non-spherical shape and several jets of the hot blue shock just emerging. This is caused by clumping and jetting at the shocked interface.

\begin{figure*}
\centering
\includegraphics[width=9cm]{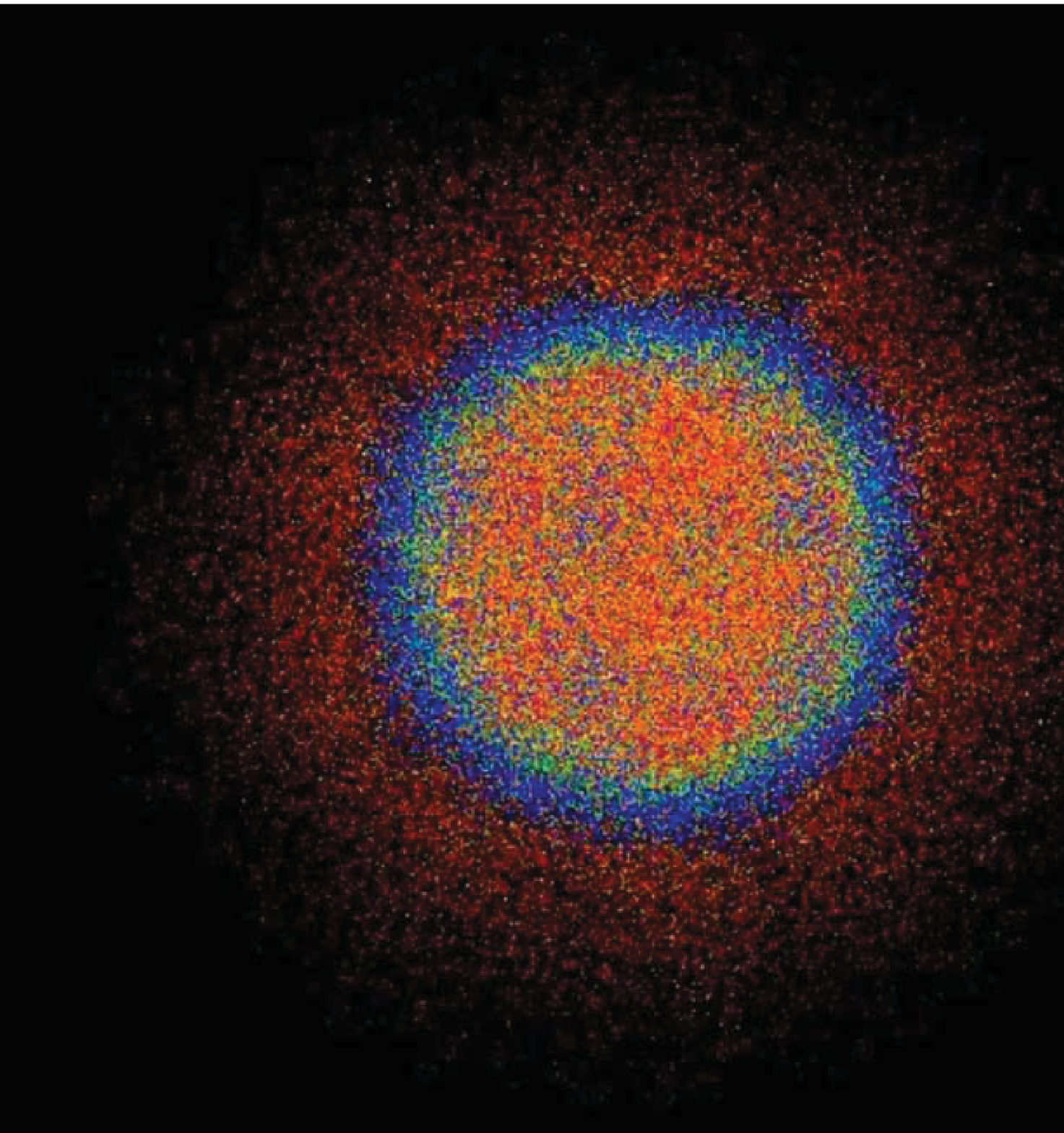}
\vskip0pt
\caption{A view of RR Lyrae at 0.8 d of model time, just as a hot shock is emerging into the extended atmosphere}
\label{o1039}
\end{figure*}

An animation showing the full motion of this model can be viewed here:

{\noindent \small \bf http://www.stellingwerf.com/rfs-bin/index.cgi?action=GetDoc\&id=135\&filenum=1}

\section{Summary}

This memo demonstrates that a 3-D hydrocode capable of modeling fluid instability and turbulence can be adapted to model a pulsating star. Two processes only vaguely understood, turbulent mixing and the dynamics of an extended atmosphere/wind seem to be modeled reasonably even with this simple setup. The next step is to extend the model to include radiation, ionization, rotation, and magnetic fields.

\end{document}